\documentclass[]{spie}  

\usepackage{amsmath,amsfonts,amssymb}
\usepackage{graphicx}
\usepackage[colorlinks=true, allcolors=blue]{hyperref}
\usepackage{upgreek}
\usepackage{epstopdf}
\usepackage{epsfig}

\usepackage{acronym}
\usepackage[caption=false]{subfig}

\acrodef{AO}[AO]{adaptive optics}
\acrodef{BS}[BS]{beamsplitter}
\acrodef{ELT}[ELT]{Extremely Large Telescope}
\acrodef{ExAO}[ExAO]{extreme adaptive optics}
\acrodef{FWHM}[FWHM]{full-width at half-maximum}
\acrodef{IFS}[IFS]{integral field spectroscopy}
\acrodef{IFU}[IFU]{integral field unit}
\acrodef{KOOL}[KOOL]{Königstuhl Observatory Opto-mechatronics Laboratory}
\acrodef{mas}[mas]{milli-arcseconds}
\acrodef{MCF}[MCF]{multi-core fiber}
\acrodef{MFD}[MFD]{mode-field diameter}
\acrodef{MMF}[MMF]{multi-mode fiber}
\acrodef{MLA}[MLA]{micro-lens array}
\acrodef{MPIA}[MPIA]{Max Plank Institute for Astronomy}
\acrodef{LED}[LED]{light-emitting diode}
\acrodef{NA}[NA]{numerical aperture}
\acrodef{NIR}[NIR]{near-infrared}
\acrodef{PGMEA}[PGMEA]{propylene-glycol-methyl-ether-acetate}
\acrodef{POP}[POP]{physical-optics propagation}
\acrodef{PSF}[PSF]{point spread function}
\acrodef{RHEA}[RHEA]{replicable high-resolution exoplanet and asteroseismology spectrograph}
\acrodef{SCExAO}[SCExAO]{Subaru Coronagraphic Extreme Adaptive Optics}
\acrodef{SEM}[SEM]{scanning electron microscopy}
\acrodef{SM}[SM]{single-mode}
\acrodef{SMF}[SMF]{single-mode fiber}
\acrodef{VSI}[VSI]{vertically-scanned white-light interferometry}

\title{An innovative integral field unit upgrade with 3D-printed micro-lenses for the RHEA at Subaru}

\author[a,b,c]{Theodoros Anagnos}
\author[d,e]{Pascal Maier}
\author[c]{Philipp Hottinger}
\author[i]{Christopher H. Betters}
\author[a,j]{Tobias Feger}
\author[i]{Sergio G. Leon-Saval}
\author[g,k]{Itandehui Gris-S\'{a}nchez}
\author[g]{Stephanos Yerolatsitis}
\author[h]{Julien Lozi}
\author[g]{Tim A. Birks}
\author[h]{Sebastian Vievard}
\author[l]{Nemanja Jovanovic}
\author[o]{Adam D. Rains}
\author[o]{Michael J. Ireland}
\author[c,p]{Robert J. Harris}
\author[a,b]{Blaise C. Kuo Tiong}
\author[h]{Olivier Guyon}
\author[i]{Barnaby Norris}
\author[m,n]{Sebastiaan Y. Haffert}
\author[d,e]{Matthias Blaicher}
\author[d,e]{Yilin Xu}
\author[q]{Moritz Straub}
\author[p]{Jörg-Uwe Pott}
\author[q]{Oliver Sawodny}
\author[q]{Philip L. Neureuther}
\author[a,b]{David W. Coutts}
\author[a,b]{Christian Schwab}
\author[d,e,f]{Christian Koos}
\author[c]{Andreas Quirrenbach}

\affil[a]{Department of Physics and Astronomy, Macquarie University, NSW 2109, Australia}
\affil[b]{MQ Photonics Research Centre, Department of Physics and Astronomy, Macquarie University, NSW 2109, Australia}
\affil[c]{Landessternwarte, Zentrum f\"ur Astronomie der Universit\"at Heidelberg, K\"onigstuhl 12, 69117 Heidelberg, Germany}
\affil[d]{Institute of Microstructure Technology (IMT), Karlsruhe Institute of Technology (KIT), Hermann-von-Helmholtz-Platz 1, 76344
Eggenstein-Leopoldshafen, Germany}
\affil[e]{Institute of Photonics and Quantum Electronics (IPQ), Karlsruhe Institute of Technology (KIT), Engesserstr. 5, 76131 Karlsruhe}
\affil[f]{Vanguard Photonics GmbH, Hermann-von-Helmholtz-Platz 1,76344 Eggenstein-Leopoldshafen, 76227 Karlsruhe}
\affil[g]{Department of Physics, University of Bath, Claverton Down, Bath, BA2 7AY, UK}
\affil[h]{National Institutes of Natural Sciences, Subaru Telescope, National Astronomical Observatory of Japan, Hilo, Hawaii, United States}
\affil[i]{University of Sydney, Sydney Institute for Astronomy, Institute for Photonics and Optical Science, School of Physics, Camperdown, Australia}
\affil[j]{Redback Systems Pty Ltd, Sydney, Australia}
\affil[k]{ITEAM Research Institute, Universitat Polit\`{e}cnica de Val\`{e}ncia, Camino de Vera, 46022 Valencia, Spain}
\affil[l]{California Institute of Technology, 1200 E. California Blvd., Pasadena CA, 91125, USA}
\affil[m]{Leiden Observatory, Leiden University, PO Box 9513, Niels Bohrweg 2, 2300 RA Leiden, The Netherlands}
\affil[n]{Steward Observatory, University of Arizona, 933 North Cherry Avenue, Tucson, Arizona}
\affil[o]{Research School of Astronomy and Astrophysics, Australian National University, Canberra,
ACT 2611, Australia}
\affil[p]{Max-Planck-Institute for Astronomy, Königstuhl 17, 69117, Heidelberg, Germany}
\affil[q]{Institute for System Dynamics, University of Stuttgart, Waldburgstr. 19, 70563 Stuttgart, Germany}

\authorinfo{Further author information: (Send correspondence to Th.A.)\\E-mail: anagnos.theodoros@gmail.com}

\begin{document} 
\maketitle

\begin{abstract}
In the new era of \acp{ELT} currently under construction,
challenging requirements drive spectrograph designs
towards techniques that efficiently use a facility’s
light collection power. Operating in the \ac{SM} regime,
close to the diffraction limit, reduces the footprint of the 
instrument compared to a conventional high-resolving
power spectrograph. The custom built injection fiber
system with 3D-printed micro-lenses on top of it
for the \ac{RHEA} at Subaru in combination with extreme 
adaptive optics of SCExAO, proved its high efficiency
in a lab environment, manifesting up to $\sim$77\% of 
the theoretical predicted performance.

\end{abstract}


\section{Introduction}
\label{sec:intro}

A wealth of crucial information can be collected through
astronomical spectroscopy, such as chemical composition,
motion parameters as well as the indirect discovery of 
celestial bodies in orbit around other stars\cite{Massey:2013}.

Conventional spectrograph designs began to make use of fibers
half a century ago\cite{Hubbard:1979,Powell:1984}
in order to enable more efficient observations, as
it became possible to locate the instrument off the telescope.
Soon after, fiber based \ac{IFU} systems were developed
\cite{Allington-Smith:2006} that allowed flexibility in 
arranging spectra on a given detector space. Early on, \acp{MMF} 
were used for the \ac{IFU}, which had high throughput for seeing-limited
starlight (e.g. Ref. \citenum{Ge:1998,Croom:2012}). 
Afterwards, new designs of \ac{IFU} systems emerged
taking advantage of \acp{SMF} (e.g. Ref.
\citenum{Leon-Saval:2012,Tamura:2012}). 

Using \acp{SMF} and operating in the diffraction limit
reduces the footprint of the instrument, however
major  limitations apply in coupling efficiency under 
seeing-limited conditions. By making use of the \ac{ExAO}
systems installed in state-of-the-art telescopes,
the coupling efficiency gets significantly better
(e.g. Ref. \citenum{Jovanovic:2017,Guyon:2020}).

While high-spatial resolution spectroscopy is achieved
by using \ac{SM}-\acp{IFU}, giving access to many
new science capabilities, the coupling losses are high 
due to the low fill fraction of \acp{SMF} and
the requirement for sub-$\upmu$m precision in alignment.

In this study, we present an upgrade of the \ac{IFU}
system on the \ac{RHEA} at Subaru \cite{Feger:2014,Rains:2016}.
This custom \ac{IFU} makes use of a \ac{MCF}
with 19 \ac{SM} cores, with 3D-printed micro-lenses
on top of the cores manufactured by the two-photon 
polymerization lithography technique 
\cite{Dietrich:2017,Hottinger:2018}. This custom
injection system significantly increases the free-space
coupling of starlight into the fiber cores while
allowing more tolerance for misalignment errors in targeting.
The \ac{IFU} system is optimized using \texttt{Zemax}
optical software for instantaneous angular sky areas
of 11 and 18\,\ac{mas} per lenslet. The system also offers a relatively high coupling efficiency and fill
factor due to the 3D-printed \ac{MLA}.

In Section \ref{sec:methods} we present the core
design and parameters, complemented by the detailed
experimental design description for the characterization
of its performance. In Section \ref{sec:results} the 
laboratory results are presented. We draw conclusions
in Section \ref{sec:conclusions} and detail our future
plans in Section \ref{sec:further_work}.

\section{Methods}
\label{sec:methods}
To increase the efficiency of light coupling from the 
8-m Subaru telescope into the \ac{IFU} feeding the
\ac{RHEA}, the following components are necessary: the 
\ac{SCExAO} system, the \ac{IFU} itself with the 3D printed
\acp{MLA}, the \ac{MCF} and the spectrograph adapted to
the output of the \ac{MCF}. Below, these components are
presented in more detail.

\subsection{Initial simulations}
\label{sec:sim-arch}

In order to simulate the intensity distribution
in the entrance of the \ac{IFU} system, all the 
optical components of the \ac{SCExAO} were taken
into account. First of all, the output beam of the
8-m Subaru telescope undergoes \ac{AO} correction
in the AO188 unit and is then routed to the visual
bench of \ac{SCExAO}. The beam intensity distribution
at this stage can be described as an Airy pattern.

\begin{figure}
\centering
\includegraphics[width=0.6\textwidth]{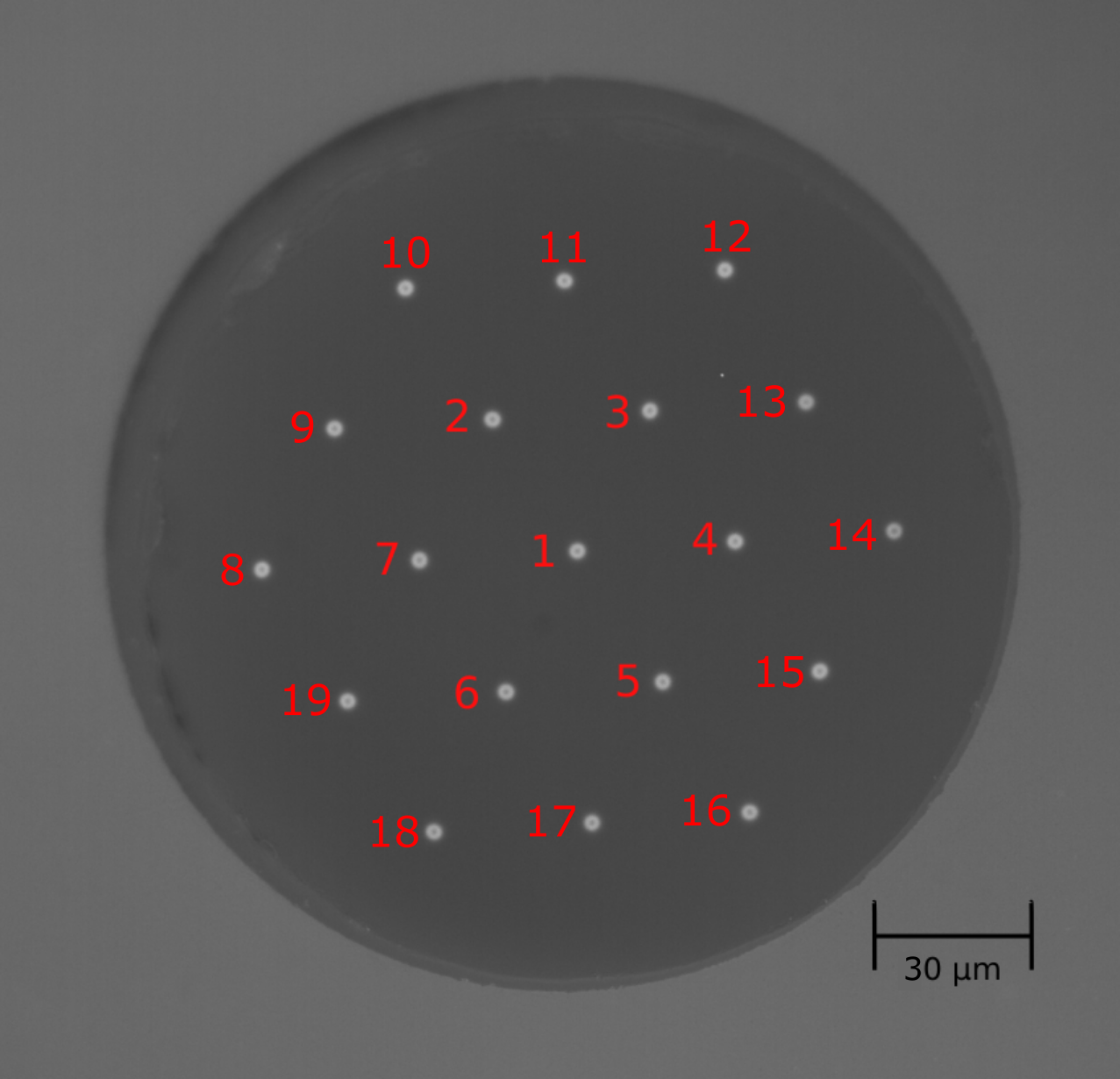}
\caption{\ac{MCF} end face after polishing and 
packaging into an FC/PC connector. The cores are
positioned in a hexagonal formation with a pitch of
30\,$\upmu$m. Each of the cores has a 6.1\,$\upmu$m
\ac{MFD} at 600\,nm ($\mathrm{1/e^{2}}$). A HeNe laser
source was used to back illuminate the fiber cores, 
which are labeled with red numbers for referencing.}
\label{fig:mcf-x-sect}
\end{figure}

For an efficient coupling of the Airy disk into
the \ac{SM} cores of the fiber, a special injection
unit is necessary. This unit is specially 
designed to match the specifications of the 
\ac{MCF}. The \ac{MCF} has a 6.1\,$\upmu$m 
\ac{MFD} at 650\,nm ($\mathrm{1/e^{2}}$)
where the \ac{SM} cut-off limit is at $\sim$600\,nm
(Figure \ref{fig:mcf-x-sect} shows a microscope
image of the \ac{MCF} end face). The cores have a
pitch of 30\,$\upmu$m in hexagonal formation, which
leads to a ratio of 4.9:1, giving enough separation 
to eliminate the cross-coupling between cores. The
3D printed structure of the \acp{MLA} is applied 
directly on top of the fiber end face using the two-photon
lithography technique described below. This
custom \ac{IFU} will be installed into \ac{RHEA} to
maximize the system potential. The structure of the \acp{MLA} was optimized for a
platescale of 11 and 18\,\ac{mas} on the sky per
lenslet given that the diffraction limit of the
Subaru telescope with the \ac{SCExAO} at 650\,nm
is 17\,\ac{mas}.

\subsection{The visual SCExAO infrastructure}
\label{sec:scexao-infr}

\begin{figure}
\centering
\includegraphics[width=0.8\textwidth]{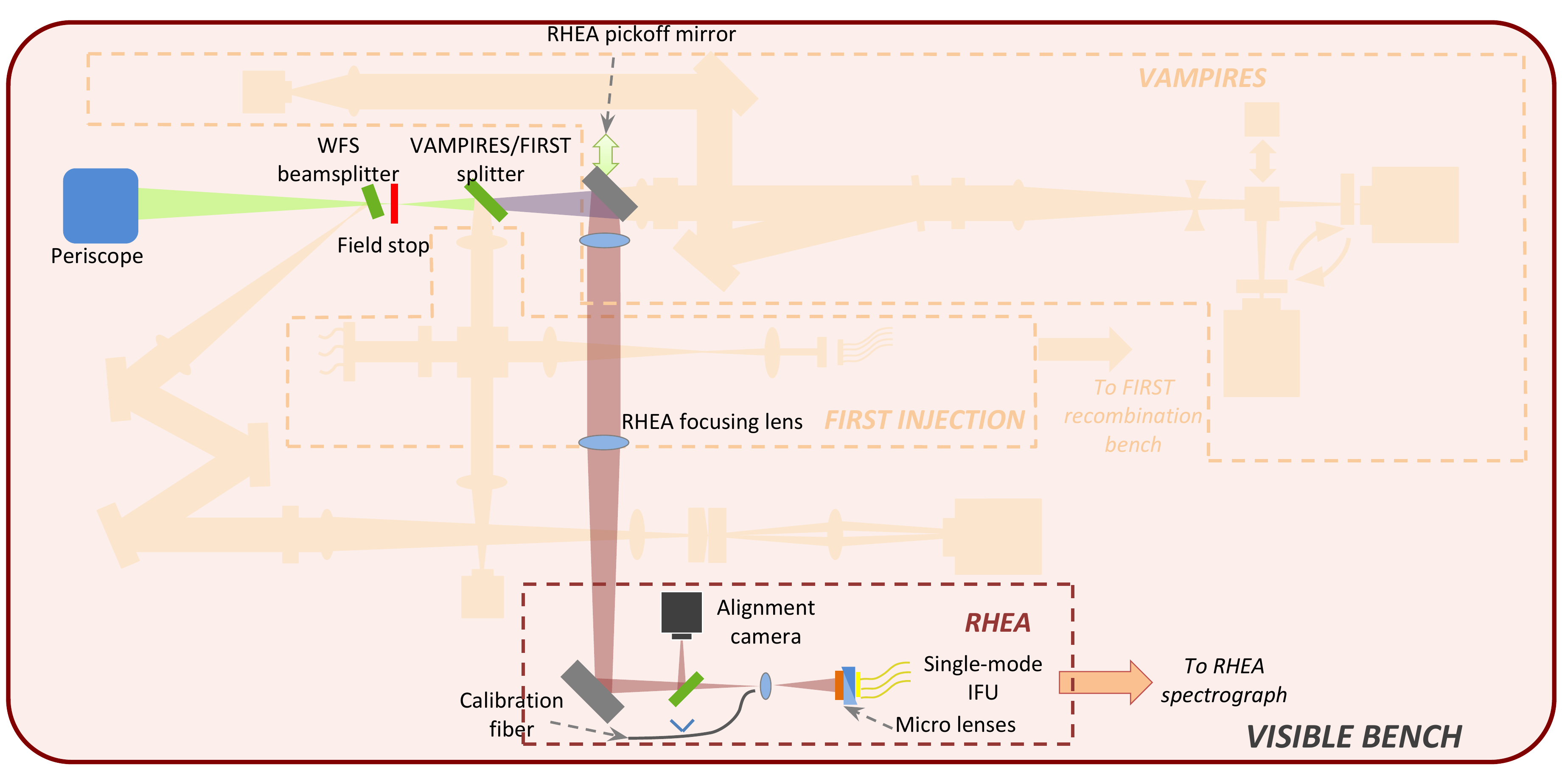}
\caption{The visual bench of the \ac{SCExAO}. The \ac{IFU} 
injection is located in the bottom center of the illustration.
The input beam from AO188 to the periscope to the \ac{IFU}
is represented with green and brown color.}
\label{fig:bench-scexao}
\end{figure}

The \ac{SCExAO} is located in the \ac{NIR} Nasmyth
focus of the 8\,m Subaru Telescope. A complete
and far more detailed description of both the
visual and \ac{NIR} paths of \ac{SCExAO} is
provided in Ref. \citenum{Jovanovic:2015}. A brief overview starts with the starlight entering the Subaru Telescope and a 30-40\% Strehl correction 
on the \ac{PSF} for the H-band accomplished by AO188 
\cite{Hayano:2008,Hayano:2010,Minowa:2010}. 
The starlight then enters the \ac{SCExAO}
\ac{NIR} bench, where the wavefront is further 
corrected for higher-order aberrations caused
by the atmosphere. In the next step, a dichroic 
filter separates the light beam in two paths:
the visible ($<$900\,nm) and the \ac{NIR}
($>$900\,nm) channel. Finally, the light beam
is focused down with optical lenses onto the 
3D printed \ac{MCF} surface (see Figure
\ref{fig:bench-scexao}).

\subsection{Throughput simulations}
\label{sec:sims}
For our simulations, an Airy disk was used an an
input to the \ac{IFU} system, taking into account
the Subaru telescope profile and the key components
of \ac{SCExAO} \cite{Jovanovic:2015,Jovanovic:2017}.

The simulation of the the Airy profile was performed
by using the \ac{POP} module of \texttt{Zemax} \cite{zemax}.
This Airy disk output feeds a performance optimizer that selects for the best \ac{MLA} structure to be 3D printed
on top of the \ac{MCF}. \ac{POP} operands were used
for varying \ac{MLA} geometries that affect the Airy beam coupling into the \ac{MCF} cores.
Several geometrical shapes were tested for the \ac{MLA}
structure in order to increase the coupling efficiency
into the cores. Finally, a spherical surface was selected,
as the alternatives provided negligible improvement in performance.
A spherical surface \ac{MLA} of 272\,$\upmu$m in height
and 115\,$\upmu$m in radius of curvature achieved the
highest coupling efficiency for the wavelength range
of 600-800\,nm, resulting in a throughput of 50\% for
the 18\,\ac{mas} platescale for the central lens and
a throughput of 21.9\% for the platescale of 11\,\ac{mas}
for the central lens. To achieve a fill factor
of $\sim$100\% in between the cores of the \ac{MCF},
the micro-lenses were merged together forming a 
hexagonal effective aperture. The roughness of the
\ac{MLA} structure is expected to be better than
$\lambda/16$-$\lambda/21$ at the working wavelength, as a 
surface roughness of 37\,nm is achieved using the
3D printing technique \cite{Dietrich:2018}.

\subsection{Micro-lens array fabrication}
\label{sec:man-proc}
The \ac{MLA} was printed to the cleaved facet of
the \ac{MCF} as a single model block using the
commercially available negative-tone photoresist
IP-Dip \cite{IP-Dip} and an in-house built 
two-photon lithography machine. This 
system is equipped with a 780\,nm femtosecond
laser \cite{Laser} and a 40x Zeiss objective lens
with \ac{NA}\,=\,1.4. A custom control software
was developed in-house to guarantee optimum 
shape-fidelity of the printed \ac{MLA} and allow
for high-precision alignment with respect to the
fiber cores of the \ac{MCF}.

\begin{figure}
\centering
\includegraphics[width=0.6\textwidth]{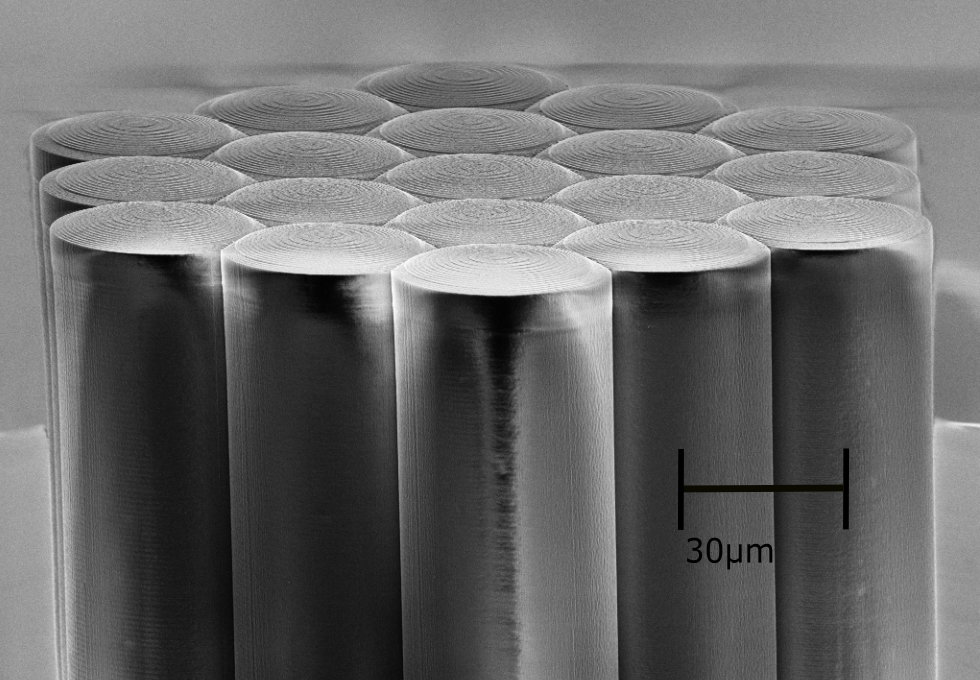}
\caption{The top structure of the 3D printed 
\ac{MLA} on top of the \ac{MCF} end face acquired
using the \ac{SEM} technique.}
\label{fig:mla-sem}
\end{figure}

As a first step, the \ac{MCF} was manually glued
to an FC-PC connector and subsequently polished to
achieve a flat fiber end-facet accessible for the
lithography machine for printing. Thereafter, the
fiber is back-illuminated by coupling in the light 
of a red \ac{LED} to accommodate machine vision for
the detection of the 19 cores of the \ac{MCF}. 
After the detection procedure, the individual 
lenses of the \ac{MLA} are aligned with respect 
to the detected core positions of the \ac{MCF}, 
thereby taking into account variations of the
core positions and pitch. All individually positioned
lenses are then merged into a single 3D-model of
the \ac{MLA} to prevent unnecessary double-illumination
in the overlap regions during the printing process.
The structure is further automatically adapted
to compensate for any tilt of the fiber end-facet.
For the purpose of reducing the required printing
time, the \ac{MLA} is divided into two parts: the
first block of the model up to just below the lens
surfaces was written with a distance between subsequent
layers, i.e., slicing distance, of 600\,nm. For
optimal printing quality, the remaining second model
block comprising the 19 lens surfaces of the \ac{MLA}
was written with a slicing distance of 100\,nm. The
writing distance between subsequent lines, i.e., 
hatching distance, was set to 100\,nm throughout the
full model. The fabricated structure was afterwards
developed in \ac{PGMEA}, flushed with isopropanol, 
and subsequently blow dried.
In the next stage, \ac{SEM} and \ac{VSI} images of 
the structure were acquired to check the quality of 
the manufacturing process (see Figure \ref{fig:mla-sem}).

\subsection{Laboratory throughput measurement setup}
\label{sec:throughput}

\begin{figure}
\centering
\includegraphics[width=0.6\textwidth]{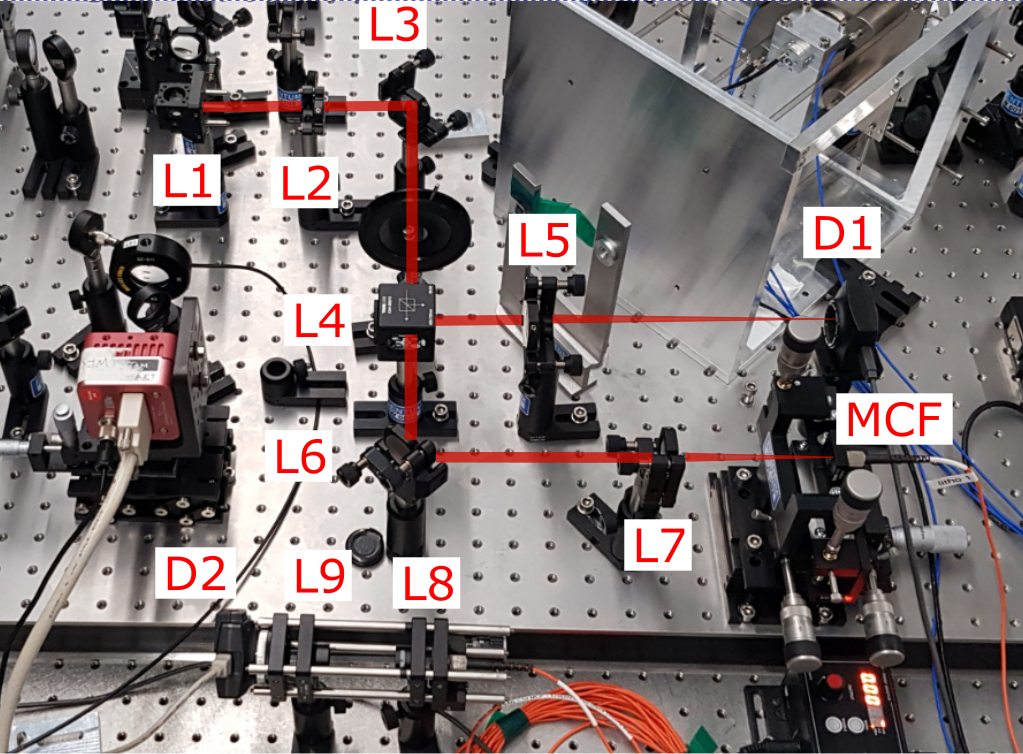}
\caption{The throughput experimental setup for 
measuring the efficiency of the custom \ac{IFU}. 
A power meter is used to calibrate the throughput. 
A set of achromat lenses  (L2-L5-L7) is used for
collimation and focusing of the beam, beamsplitter
(BS) and CMOS detector (D1, D2) for imaging the 
near-field output of the \ac{MCF} (L8, L9).}
\label{fig:th_setup}
\end{figure}

For measuring the throughput performance of the 
custom 3D-printed \ac{MCF}, a set of opto-mechanical
parts was constructed as an addition to the \ac{KOOL} test-bed.
The throughput setup is presented in Figure 
\ref{fig:th_setup}. The HeNe laser light (632\,nm) passes
through a 50:50 non-polarizing \ac{BS} L1 (Thorlabs
CM1-BS014) and is collimated using an achromat 
lens L2 (Thorlabs AC127-025-B-ML). Later on, the beam 
is split using another 50:50 non-polarizing \ac{BS}
L4 (Thorlabs CM1-BS014). One beam is focused down
using an 100\,mm achromat (AC254-100-B-ML) to the
CMOS detector (Thorlabs DCC1545M), and the other
is routed through a flip mirror L6 to an achromat
L7 (AC254-060-B-ML or AC254-100-B-ML depending on the 
platescale) and focused down to the 3D-printed \ac{MCF}
that is mounted onto a 4-axis mount (Thorlabs MBT401D).
After that, the fiber exit is re-imaged using a
combination of achromats L8 and L9 (AC127-019-B-ML
and AC127-050-B-ML) to the CMOS detector
(Thorlabs DCC1545M). To calculate the total
throughput of the \ac{MCF} including the the
coupling losses, a power meter was used (Thorlabs 
S120C) in order to perform the measurements and 
calibrate absolute flux through the re-imaging system. 

\section{Results}
\label{sec:results}

\subsection{Throughput efficiency results}
To asses the performance of the custom \ac{IFU}
system before installation into \ac{RHEA},
laboratory tests were performed as described in 
section \ref{sec:throughput}. The outcome of these
tests are presented here.

As mentioned in section \ref{sec:throughput} the 
throughput measurements were monochromatic using a 
HeNe laser at 632\,nm. After a series of optical
elements in the \ac{KOOL} test-bed, the beam was focused
down to the 3D-printed \ac{MCF}. Data frames with
exposure times of a fraction of a second were collected
with both achromats L7 (AC254-060-B-ML or AC254-100-B-ML),
using the setup described above. The setup was able
to sample the near-field of the \ac{MCF} exit and 
filter the light between the adjacent cores
of the fiber. Averaged dark data frames were recorded
as well, and subtracted from the data for further 
processing.

To characterize the absolute performance of the \ac{IFU},
two separate experiments were conducted; in the first,
the total coupling efficiency of the light into each
of the 19 cores was determined after the alignment
of the cores on-axis with the injected beam. In the 
second experiment, the misalignment tolerances 
of the injected beam were measured by translating
the injected beam by steps of 5\,$\upmu$m in respect 
to the central core of the \ac{MCF}. This was more representative
of realistic on-sky conditions where the star 
would be moving due to atmospheric perturbations.

\begin{figure}
\centering
\includegraphics[width=\textwidth]{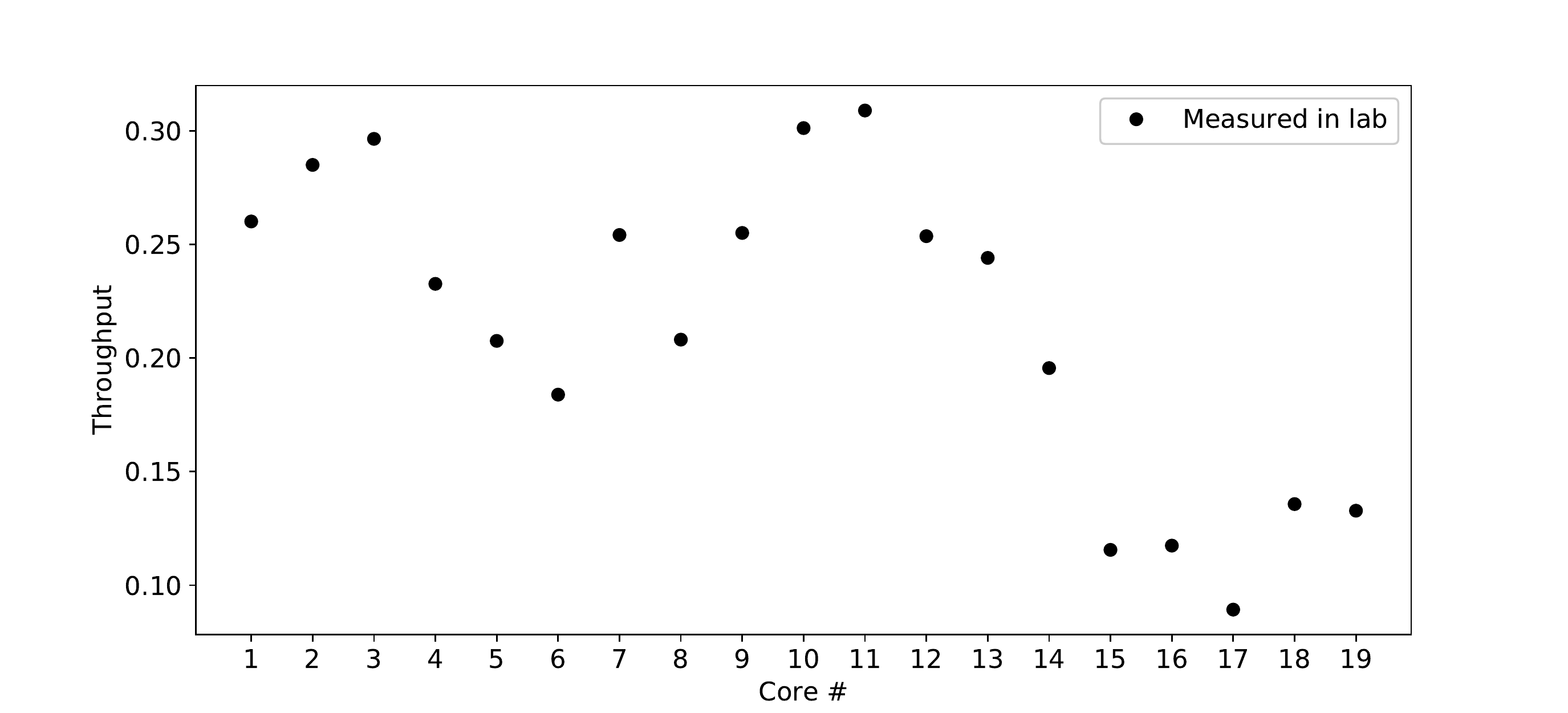} 
\caption{Coupling efficiency for all of the cores
of the \ac{MCF} (see Figure \ref{fig:mcf-x-sect} for
the numbering of the cores). This is shown for the 
18\,\ac{mas} platescale.}
\label{fig:th-cores}
\end{figure}

The results of the first experiment for the 18\,\ac{mas} 
platescale are shown in Figure \ref{fig:th-cores}. The
average coupling efficiency was 21.45\,$\pm$\,3\% with
a maximum of 30.89\,$\pm$\,3\% for the core \#11. 
The maximum measured throughput corresponds to
77.41\% of the simulated value (40\%) calculated 
with \texttt{Zemax}. For the 11\,\ac{mas} 
platescale, the coupling efficiency of the 
central core was 16.9\,$\pm$\,3\% (77.2\% of the 
simulated value). The total throughput from all of 
the cores summed was calculated, representative
of an unresolved target, and was measured  
to be 41.5\,$\pm$\,2\%. The residual coupling losses are
associated with the imperfectly polished fiber, 
Fresnel reflections ($\sim$4\%), and mode-field 
mismatch at the focus.

\begin{figure}
\centering
\includegraphics[width=0.6\textwidth]{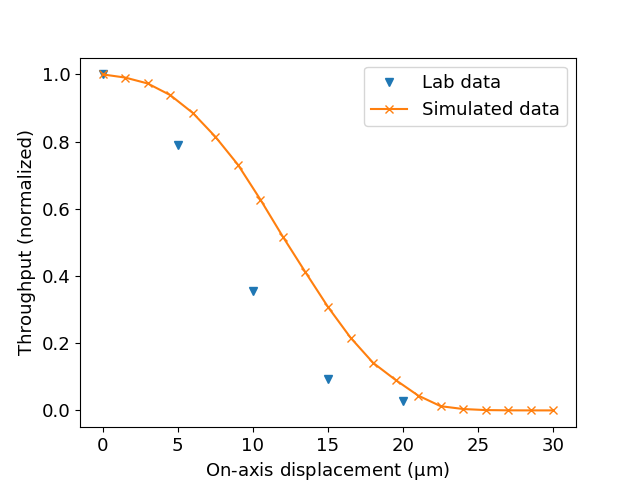} 
\caption{Coupling efficiency of the central
core of the \ac{MCF} as a function of off-axis
target, compared with the simulated data 
from \texttt{Zemax}. Results are normalized 
to the maximum coupling efficiency including the 
errors (smaller than the data points).}
\label{fig:th-latt}
\end{figure}

In Figure \ref{fig:th-latt} the results from the 
second experiment are presented. This illustration shows the
coupling tolerances as a function of off-axis
target injection for the case of 18\,\ac{mas}.
This demonstrates the potential of 3D-printed \ac{MLA}
technology for misaligned targets, showing 
fairly good coupling efficiency even for an 
off-axis injection of $\sim$10\,$\upmu$m, retaining
$\sim$40\% of the maximum throughput.

\begin{figure}
\centering
\includegraphics[width=\textwidth]{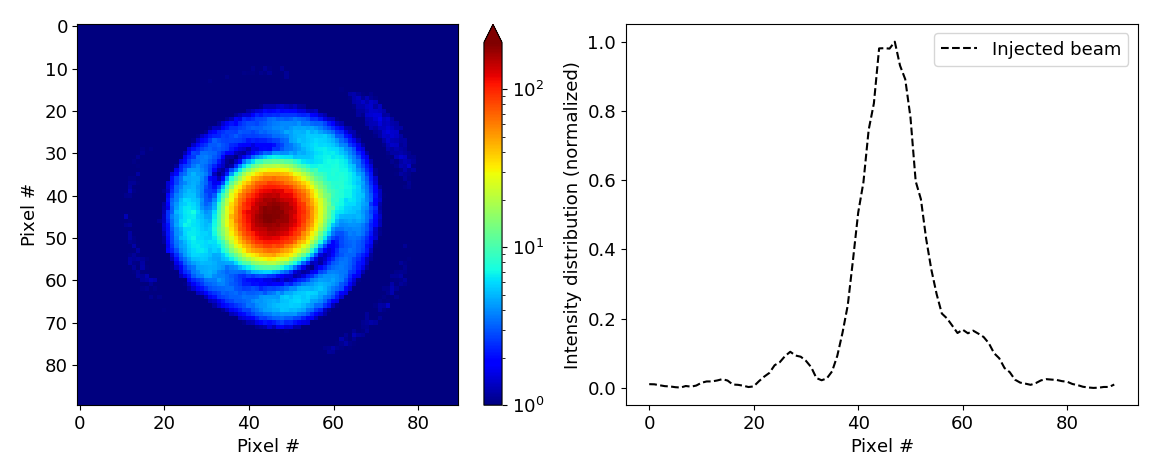} 
\caption {\textbf{Left panel:} 2D image of the
injected beam in logarithmic color scale for
better clarity, as measured in the laboratory.  
\textbf{Right panel}: Intensity profile of the 
injected point spread function normalized to its
maximum, from the 2D image data.}
\label{fig:inj_psf}
\end{figure}

\section{Conclusions}
\label{sec:conclusions}

In this work we presented a novel \ac{IFU} system
upgrade for \ac{RHEA} at the Subaru telescope. This
\ac{IFU} is composed of a custom \ac{MCF} with
3D printed micro-lenses on top of the cores to
increase the coupling efficiency for off and
on-axis targets from \ac{SCExAO} at the Subaru 
8\,m telescope. The \ac{IFU} system is optimized
using the \texttt{Zemax} \ac{POP} module for an 
on-sky angular dimension of 11 and 18\,\ac{mas}
using an Airy profile beam produced by the visual
arm of \ac{SCExAO}.

The custom \ac{MCF} is composed of 19 cores in the 
same cladding with a core-to-core spacing of 
30\,$\upmu$m and a 6.1\,$\upmu$m \ac{MFD} at
650\,nm ($\mathrm{1/e^{2}}$) which leads to a 
negligible cross-coupling between the cores. The
cores are positioned in a hexagonal formation and
their cut-off \ac{SM} limit is above 600\,nm.

The structure of the \ac{MLA} was manufactured 
with two-photon lithography and 3D-printed
on top of the cores of the \ac{MCF}, significantly
enhancing the throughput of light into the fiber cores
from levels of few percent to a maximum of
30.89\,$\pm$\,3\% for on-axis targets for a platescale of
18\,\ac{mas}. Furthermore, the custom \ac{MLA}
improved the off-axis light losses even for a
10\,$\upmu$m lateral injection. The throughput
performance across all the \ac{MLA} as a representative
of a single unresolved target was 41.5\,$\pm$\,2\%.
The laboratory results correspond to $\sim$77\%
of the simulated results. The difference in throughput
performance from simulated results are likely 
associated with the imperfectly polished \ac{MCF},
Fresnel reflections ($\sim$4\%) and mode-field
mismatch at the focus. These lab results were
performed at the \ac{KOOL} test-bed with a HeNe
laser source (632\,nm).

\section{Further work}
\label{sec:further_work}
Plans for further work include a separate 3D-printed
\ac{MCF}, optimized for only the 18\,\ac{mas} platescale
with \texttt{Zemax} simulations. Both \ac{IFU}
systems will be tested on the \ac{KOOL} test bed
including the effect of atmospheric turbulence
using the \ac{AO} system of the \ac{KOOL} 
infrastructure.

Future work will be to integrate the fibers 
into the \ac{RHEA} and perform on-sky tests on 
a variety of targets (resolved, un-resolved stars,
confirmed exoplanets, spectroscopic standard stars
and double star systems) in order to probe its 
scientific potential.

\acknowledgments
T.A. is a fellow of the International Max Planck
Research School for Astronomy and Cosmic Physics at
the University of Heidelberg (IMPRS-HD) and is 
supported by the Cotutelle International Macquarie 
University Research Excellence Scholarship. P.M., M.B.,
Y.X. and C.K. are supported by Bundesministerium f\"{u}r 
Bildung und Forschung (BMBF), joint project PRIMA
(13N14630), the Helmholtz International Research 
School for Teratronics (HIRST), Deutsche 
Forschungsgemeinschaft (DFG, German Research Foundation)
under Germany's Excellence Strategy via the Excellence 
Cluster 3D Matter Made to Order (EXC2082/1-390761711).
R. J. H. and P.H. are supported by the Deutsche Forschungsgemeinschaft
(DFG) through project 326946494, 'Novel Astronomical 
Instrumentation through photonic Reformatting'.
T.B. \& S.Y. are supported from the European Union's
Horizon 2020 grant 730890, and from the UK Science and
Technology Facilities Council grant ST/N000544/1.
S.Y.H. is supported by the NASA Hubble Fellowship grant
\#HST-HF2-51436.001-A awarded by the Space Telescope
Science Institute, which is operated by the Association 
of Universities for Research in Astronomy, Incorporated,
under NASA contract NAS5-26555.
The development of SCExAO was supported by the JSPS 
(Grant-in-Aid for Research \#23340051, \#26220704 
\#23103002), the Astrobiology Center (ABC) of the 
National Institutes of Natural Sciences, Japan, the
Mt Cuba Foundation and the directors contingency
fund at Subaru Telescope, and the OptoFab node of 
the Australian National Fabrication Facility. The
authors wish to recognize and acknowledge the very 
significant cultural role and reverence that the
summit of Maunakea has always had within the
indigenous Hawaiian community. We are most fortunate
to have the opportunity to conduct observations from
this mountain. This research made use of Astropy, a
community-developed core \texttt{Python} package for
Astronomy \cite{astropy:2013, astropy:2018}, Numpy
\cite{numpy} and Matplotlib \cite{matplotlib}.
Furthermore, this publication makes use of data generated at the Königstuhl Observatory Opto-mechatronics Laboratory (KOOL) which is run at the Max-Planck-Institute for Astronomy (MPIA, PI Jörg-Uwe Pott, jpott@mpia.de) in Heidelberg, Germany. KOOL is a joint project of the MPIA, the Landessternwarte Königstuhl (LSW, Univ. Heidelberg, Co-I Philipp Hottinger), and the Institute for System Dynamics (ISYS, Univ. Stuttgart, Co-I Prof. Oliver Sawodny). KOOL is partly supported by the German Federal Ministry of Education and Research (BMBF) via individual project grants.

\bibliography{references}
\bibliographystyle{spiebib}

\end{document}